\documentclass[11pt]{article}
\usepackage{moriond,epsfig}
\usepackage{amsfonts}
\usepackage{amsmath}
\usepackage{amssymb}

\def\Journal#1#2#3#4{{#1} {\bf #2}, #3 (#4)}

\def\NPB{{\em Nucl. Phys.} B}
\def\PLB{{\em Phys. Lett.}  B}

\def\PRD{{\em Phys. Rev.} D}

\def\be{\begin{equation}}
\def\ee{\end{equation}}
\def\bea{\begin{eqnarray}}
\def\eea{\end{eqnarray}}

\begin{document}
\vspace*{4cm}
\title{HIGGS BOSON DECAY RATE INTO GLUONS IN A 5 DIMENSIONAL CALCULABLE MODEL}

\author{ G. CACCIAPAGLIA }

\address{Scuola Normale Superiore and INFN, piazza dei Cavalieri 7,\\
I-56126, Italy}

\maketitle\abstracts{In an extension of the Standard Model with one compact extra dimension and N=1 supersymmetry, we compute the Higgs boson decay width into two gluons, relevant to the Higgs production in hadronic collisions. We find that at one-loop the decay width is significantly suppressed with respect to the SM. In particular, for a compactification radius $R=(370 \pm 70 GeV)^{-1}$ and a Higgs mass $m_H = 127 \pm 8 GeV$ we find it to be less than 15\% of the SM result.
}

Since a lot of time, the idea that extra spatial dimensions can exist has been very intriguing.
But, only recently~\cite{anton} it has been realized that compact dimensions can be large and in the reach of next generation experiments~\cite{wrochna}.
The main drawback of this idea, from a phenomenological point of view, is the lack of a quantitative connection between any known physical scale and the scale where such a new physics could take place.

On the other hand, the model~\cite{BHN} we consider yields a framework where this problem is avoided for a number of quantities.
The model is embedded in a 5 dimensional spacetime with $N=1$ supersymmetry and minimal matter and gauge content with respect to the Standard Model.
In particular only one Higgs hypermultiplet is present.
The extra dimension is conpactified on $\mathbb{R}^1/\mathbb{Z}_{2} \times \mathbb{Z}'_{2}$, the only orbifold that leaves only the SM fields as zero modes~\cite{BHNclas}.
Notwithstanding the presence of a Fayet-Iliopoulos term~\cite{nilles} on the branes, the electroweak symmetry breaking is radiatively driven and leads to a prediction of the Higgs mass in the range $120\div 170 GeV$. 
On the other hand, the compactification radius $R$, related to $m_H$, can increase up to a $TeV$~\cite{BHNfi}.
The relation between the FI term and possible hypercharge anomalies has also been studied~\cite{SSSZ,anom,BCCRS}.
In this framework, the presence of a residual local supersymmetry and of the gauge symmetry provides several observables with calculability.
This means that observables, such as the $\mathcal{BR} (b\rightarrow s \gamma)$~\cite{bsg}, the muon $g-2$~\cite{gmen2} and the parameter $\epsilon_3$, are insensitive to the natural cut-off of the theory, that is approximately $5/R\approx 2 TeV$.

In this proceeding we present the calculation of the 1-loop contribution of the top KK modes to the Higgs boson decay width into two gluons~\cite{Hgg}.
The relevance of this process is that it can be related to the gluon fusion production rate of the Higgs, that in the predicted range of mass is expected to be the main channel in the SM~\cite{spira}.
As the full expression of the mass eigenstates as functions of the Higgs field is available~\cite{BHN}, the calculation can be performed with high accuracy and control on the next order terms in the $R$ expansion.
The ratio between our calculation and the SM decay rate is plotted in the figure, where the full result (dashed line) for $m_H=127\: GeV$ and a particular limit ($m_H \rightarrow 0$) are shown. 
For more details we refer the reader to the letter~\cite{Hgg}.

\parbox{8.2cm}{\psfig{figure=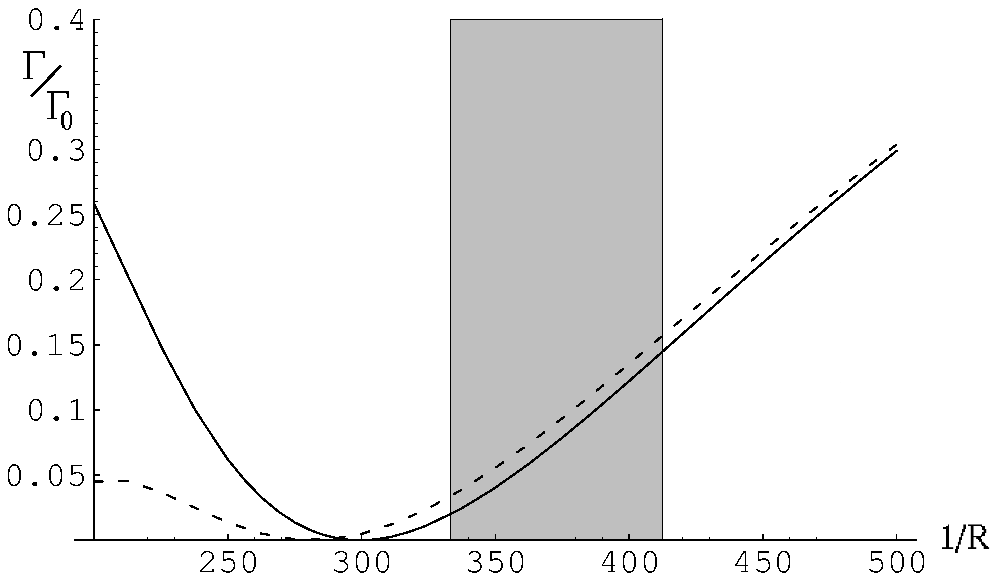,height=2in}}
\parbox{7cm}{For the value $1/R = 370 \pm 70 \: GeV$ (grey band), related to the Higgs mass $m_H = 127 \pm 8\: GeV$, preferred in the case of radiatively generated FI term~\cite{BHNfi}, the 1-loop decay rate is reduced below the 15\% level of the SM prediction.
Also taking into account that the 2-loop amplitude~\cite{2loops} is at 40\% level of the 1-loop one, we can conclude that the Higgs decay rate into two gluons is highly suppressed.}
\vspace{0.1cm}

This result shows how the Higgs phenomenology is sensitive to the presence of extra dimensions, as also pointed out in a similar context~\cite{petr}, where an enhancement was found instead.

\section*{Acknowledgments}
This work was supported by the EC grant under the RTN contract HPRN-CT-2000-00148.
I would like to thank M.Cirelli and G.Cristadoro whose collaboration this work was done with and P.Slavich for stimulating discussions during the Conference. 
Moreover I thank the organizers for the inspiring atmosphere and the amazing ski resort.

\end{document}